\documentclass[%
 aip,jcp,
 amsmath,amssymb,
 reprint,
]{revtex4-2}

\usepackage{graphicx}
\usepackage{dcolumn}
\usepackage{bm}

\usepackage[T1]{fontenc}
\usepackage[utf8]{inputenc}
\usepackage{mathptmx}
\usepackage{listings}
\usepackage{rotating} 

\usepackage{color}
\usepackage{dcolumn} 
\usepackage{bm} 
\usepackage{graphicx}
\usepackage{multirow} 
\usepackage{pifont} 
\usepackage{epsfig}
\usepackage{amsmath} 
\usepackage{subfigure}
\usepackage{float}
\usepackage{booktabs}
\usepackage{tabularx}
\usepackage{natbib}
\usepackage{gensymb}
\usepackage{rotating}
\usepackage{threeparttable}
\usepackage{comment}
\usepackage[normalem]{ulem}
\usepackage[hidelinks]{hyperref} 

\begin{document}
  
\author{Marcus Dante Liebenthal}
\affiliation{
             Department of Chemistry and Biochemistry,
             Florida State University,
             Tallahassee, FL 32306-4390}             
             
\author{A. Eugene DePrince III}
\email{adeprince@fsu.edu}
\affiliation{
             Department of Chemistry and Biochemistry,
             Florida State University,
             Tallahassee, FL 32306-4390}

\title{The Orientation Dependence of Cavity-Modified Chemistry}

\begin{abstract}

Recent theoretical studies have explored how ultra-strong light--matter coupling can be used as a handle to control chemical transformations. {\em Ab initio} cavity quantum electrodynamics (QED) calculations demonstrate that large changes to reaction energies or barrier heights can be realized by coupling electronic degrees of freedom to vacuum fluctuations associated with an optical cavity mode, provided that large enough coupling strengths can be achieved. In many cases, the cavity effects display a pronounced orientational dependence. In this Perspective, we highlight the critical role that geometry relaxation can play in such studies. As an example, we consider recent work [Nat.~Commun.~{\bf 14}, 2766 (2023)] that explored the influence of an optical cavity on Diels-Alder cycloaddition reactions and reported large changes to reaction enthalpies and barrier heights, as well as the observation that changes in orientation can inhibit the reaction or select for one reaction product or another. Those calculations used fixed molecular geometries optimized in the absence of the cavity and fixed relative orientations of the molecules and the cavity mode polarization axis. Here, we show that, when given a chance to relax in the presence of the cavity, the molecular species reorient in a way that eliminates the orientational dependence.  Moreover, in this case, we find that qualitatively different conclusions regarding the impact of the cavity on the thermodynamics of the reaction can be drawn from calculations that consider relaxed versus unrelaxed molecular structures.

\end{abstract}

\maketitle

\section{Introduction}
\label{SEC:INTRODUCTION}

Strong interactions between light and matter can induce nontrivial changes in molecular properties and chemical reactivity. Such interactions have been leveraged experimentally as a means of augmenting material properties\cite{Whittaker98_6697,Mugnier04_036404,Ebbesen15_1123,Ebbesen16_2403,Bellessa19_173902} and controlling chemical transformations carried out within optical cavities. \cite{Ebbesen12_1592, Ebbesen16_11462, Ebbesen19_615,Borjesson18_2273, George19_10635, Jeffrey22_429} Given these exciting prospects, substantial computational and theoretical effort has been undertaken to develop both effective theories\cite{YuenZhou18_6325} for simulating collective strong coupling effects and quantum electrodynamics generalizations of familiar {\em ab initio} quantum chemistry methods\cite{DePrince23_041301} for modeling cavity-induced changes to electronic structure in the single-molecule strong coupling limit. In the latter category, the last decade has seen the development and application of QED generalizations of many quantum chemistry methods, including density functional theory (QEDFT\cite{Bauer11_042107,Tokatly13_233001,Rubio14_012508,Rubio15_093001,Rubio18_992,Appel19_225,Rubio19_2757,Narang20_094116,Rubio22_7817, Rubio23_11191} or QED-DFT\cite{DePrince22_9303,Rubio22_094101,DePrince23_5264, Shao21_064107, Shao22_124104}) configuration interaction (QED-CI),\cite{Koch21_094113, Foley22_154103,Foley24_1214,Wilson24_094111} coupled-cluster theory (QED-CC),\cite{Koch20_041043, Manby20_023262, Corni21_6664, DePrince21_094112, Flick21_9100, Koch21_094113, Koch22_234103, Flick22_4995, Rubio22_094101, Knowles22_204119, DePrince22_054105, Rubio23_2766, Rubio23_10184, Koch23_4938, Koch23_8988, Koch23_031002, DePrince23_5264} and more.\cite{DePrince22_053710, Dreuw23_124128,Narang23_arXiv:2307.14822}

Many {\em ab initio} QED studies have focused on ground-state effects, exploring how vacuum fluctuations modify ground-state electronic structure and how these changes can be leveraged for useful purposes in chemistry applications. Within these studies, a common theme emerges: ground-state cavity effects have a pronounced orientational dependence. Specific examples using various {\em ab initio} QED approaches include the following. Vu, McLeod, Hanson, and DePrince\cite{DePrince22_9303} applied time-dependent QED-DFT to model cavity-embedded BINOL ([1,1'-binaphthalene]-2,2'-diol) derivatives in the context of the enantiopurification of BINOL via chiral-group directed photoisomerization.\cite{Hanson19_1263} This work found that the diestereometric excess predicted in the absence of the cavity could be enhanced, suppressed, or even inverted, depending on the relative orientation of the molecule and the cavity-mode polarization axis. Pavo{\v{s}}evi{\'{c}}, Hammes-Schiffer, Rubio, and Flick\cite{Flick22_4995} used QED-CC to study cavity effects on proton transfer reactions in malonaldehyde and aminopropenal. This work found that the reaction barrier height can increase or decrease depending on the molecule's orientation in the cavity.  Pavo{\v{s}}evi{\'{c}}, Smith, and Rubio demonstrated using QED-CC that similar enhancement / inhibition could be achieved in Diels-Alder reactions of cyclopentadiene and acrylonitrile\cite{Rubio23_2766} and azide-alkyne cycloaddition reactions.\cite{Rubio23_10184} These studies also showed that orientational control could be used to selectively steer the system toward a specific reaction product. Severi and Zerbetto\cite{Zerbetto23_9145} used a QED generalization of Hartree-Fock theory to show that the barrier to isomerization in butadiene can be enhanced or suppressed depending on the relative orientations of the molecule and cavity mode polarization axis. Haugland, Sch{\"a}fer, Ronca, Rubio, and Koch\cite{Koch21_094113} have also shown using QED-CC and QED full CI that intermolecular interactions can be modulated in a way that stabilizes or destabilizes non-covalently bound complexes, depending on how they are oriented in the cavity.

These studies all suggest that subtle quantum electrodynamical effects can be used to dictate chemical outcomes, assuming sufficiently strong electron-photon coupling can be achieved. Equally exciting is the level of control over selectivity, etc., that these studies imply can be achieved, given absolute control over molecular orientation. For gas or solution-phase reactions, however, such control is not necessarily possible, and, yet, essentially all {\em ab initio} QED studies assume (i) fixed orientations relative to the cavity mode polarization axis and (ii) fixed molecular geometries that are not optimized to account for any cavity interactions. Relatively little effort has gone into understanding how molecular geometries can be perturbed by cavity forces and whether claims of orientation-based selectivity are robust to such geometry relaxation effects. As a result, the literature may overstate the prospects cavity-based control over ground-state chemistry arising from electronic strong coupling. 

\begin{figure}[htbp]
 \centering
 \includegraphics[width=\linewidth]{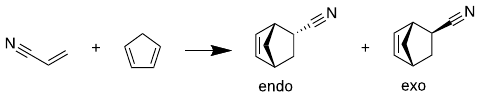}
 \caption{The Diels-Alder cycloaddition of cyclopentadiene and acrylonitrile, forming either the \textit{endo} or \textit{exo} diastereomer.} \label{fig:reaction_diagram}
\end{figure}

In this work, we consider the importance of cavity-induced geometry relaxation effects in computational studies of cavity-modified chemistry. As a model system, we choose to revisit the Diels-Alder cycloaddition of cavity-embedded cyclopentadiene and acrylonitrile that was studied by Pavo{\v{s}}evi{\'{c}}, Smith, and Rubio in Ref.~\citenum{Rubio23_2766} (see Fig.~\ref{fig:reaction_diagram}). We develop analytic energy gradients for the QED-DFT approach, which we use to optimize structures of the relevant cavity-embedded species in order to provide a side-by-side comparison of thermodynamic properties inferred from unperturbed geometries and perturbed geometries obtained at various initial relative orientations of the molecular species and cavity mode polarization axis. In short, without some physical mechanism to fix the molecular orientation, the system will reorient in a way that eliminates the orientational dependence of the reaction barrier heights and enthalpies for this cycloaddition reaction.

\section{Theory}

\label{SEC:THEORY}

\newcommand{\bra}[1]{\left\langle #1 \right|}
\newcommand{\ket}[1]{\left| #1 \right\rangle}
\newcommand{\Hpf}{\hat{H}_{\rm CS}}
\newcommand{\He}{\hat{H}_\text{e}}
\newcommand{\wCav}{\omega_{\rm cav}}
\newcommand{\bDag}{\hat{b}^\dagger}
\newcommand{\bOp}{\hat{b}}
\newcommand{\lamVec}{\bm{\lambda}}

\newcommand{\muTot}{\bm{\hat{\mu}}}
\newcommand{\muElec}{\muTot_\text{e}}
\newcommand{\muAvg}{\left\langle {\muTot} \right\rangle}
\newcommand{\muAvge}{\left\langle {\muElec} \right\rangle}
\newcommand{\muDiffe}{\lamVec \cdot \left[\muElec - \muAvge\right]}
\newcommand{\muDiff}{\lamVec \cdot \left[\muTot - \muAvg\right]}

The physics of a strongly-coupled light--matter system is captured by the Pauli-Fierz (PF) Hamiltonian,\cite{Spohn04_book,Rubio18_0118} 
which can be derived from a minimal coupling or ``${\rm p} \cdot {\rm A}$'' Hamiltonian.\cite{Huo20_6321} The PF Hamiltonian is obtained via a transformation of the ${\rm p} \cdot {\rm A}$ Hamiltonian to the length gauge ({\em i.e.}, using the Power-Zienau-Woole transformation), followed by a phase transformation and the use of the dipole approximation. Following some straightforward but tedious manipulations, one eventually arrives at
\begin{eqnarray}
    \label{EQN:PHF}
    \hat{H}_{\rm PF} &=& \He + \wCav\bDag\bOp - \sqrt{\frac{\wCav}{2}} \lamVec \cdot \muTot \left(\bDag + \bOp\right) \nonumber \\
    &+& \frac{1}{2} \left( \lamVec \cdot \muTot \right)^2
\end{eqnarray}
Here, the first two terms ($\He$ and  $\wCav\bDag\bOp$) represent the electronic and photon Hamiltonians, respectively, and $\wCav$ is the cavity mode frequency. The symbols $\hat{b}^\dagger$ and $\hat{b}$ represent photon creation and annihilation operators, respectively. The third and fourth terms are the bilinear coupling and dipole self-energy terms, respectively. In these terms, we find the molecular dipole operator, $\muTot$, and the coupling strength vector, $\lamVec$, which is related to the effective cavity mode volume ($V_{\rm eff}$) and is oriented along the transverse polarization vector, $\hat{\bm e}$:
\begin{equation}
    \label{EQN:lambda}
    \lamVec = \lambda \hat{\bm e} = \sqrt{\frac{4\pi}{V_{\rm eff}}} \hat{\bm e}
\end{equation}
The Hamiltonian in Eq.~\ref{EQN:PHF} describes coupling between electronic degrees of freedom and a single cavity mode, but it can easily be generalized to account for multiple modes. 

\subsection{QED-DFT Ground-State Energy}

To model the ground-state of $\hat{H}_{\rm PF}$, we use a quantum-electrodynamics generalization of density functional theory (DFT)  that resembles conventional Kohn-Sham DFT,\cite{Kohn64_B864} augmented by a mean-field description of electron-photon interactions. We begin by considering a mean-field solution to the entire polaritonic problem ({\em i.e.}, QED Hartree-Fock, or QED-HF), in which case the wave function for the system has the form 
\begin{equation}
\label{EQN:QED_HF}
    \ket{\Phi_{0}} = \ket{0^\text{e}} \otimes \ket{0^{\rm p}}
\end{equation}
Here, $\ket{0^\text{e}}$ is a Slater determinant of electronic orbitals and $\ket{0^{\rm p}}$ is a zero-photon state.
As described in Ref.~\citenum{Koch20_041043}, it is convenient to define the photon part of $\ket{\Phi_{0}}$ as
\begin{equation}
    \ket{0^{\rm p}} = \hat{U}_{\rm CS} \ket{0}
\end{equation}
where $\ket{0}$ is the photon vacuum state, and $\hat{U}_{\rm CS}$ is a unitary coherent-state transformation operator 
\begin{equation}
\label{EQN:U_CS}
    \hat{U}_{\rm CS} = {\rm exp}\left( z(\hat{b}^{\dagger} - \hat{b}) \right) 
\end{equation}
with
\begin{equation}\label{EQN:z}
    z = \frac{-{\bm \lambda} \cdot \langle {\muTot} \rangle }{\sqrt{2 \omega_{\rm cav}}}
\end{equation}
The expectation value of $\hat{H}_{\rm PF}$ with respect to $\ket{\Phi_{0}}$ is
\begin{align}
E_{\rm QED-HF} = \bra{\Phi_{0}} \hat{H}_{\rm PF} \ket{\Phi_{0}} =&  \bra{0^\text{e}} \otimes \bra{0^{\rm p}} \hat{H}_{\rm PF} \ket{0^{\rm p}} \otimes \ket{0^\text{e}} \nonumber \\
=& \bra{0^\text{e}} \otimes \bra{0} \hat{U}_{\rm CS}^\dagger \hat{H}_{\rm PF} \hat{U}_{\rm CS} \ket{0} \otimes \ket{0^\text{e}}\nonumber \\
=& \bra{0^\text{e}} \otimes \bra{0} \hat{H}_{\rm CS} \ket{0} \otimes \ket{0^\text{e}}
\end{align}
where the coherent-state transformed Hamiltonian, $\hat{H}_{\rm CS}$, is
\begin{eqnarray}
    \label{EQN:PFH_COHERENT}
    \Hpf &=& \He + \wCav\bDag\bOp - \sqrt{\frac{\wCav}{2}} \muDiff \left(\bDag + \bOp\right) \nonumber \\
    &+& \frac{1}{2} \left(\muDiff\right)^2
\end{eqnarray}
We can see now that the mean-field energy is simply
\begin{equation}
    E_{\rm QED-HF} = \bra{0^\text{e}} \hat{H}_\text{e} \ket{0^\text{e}} + \frac{1}{2} \bra{0^\text{e}} \left(\muDiffe\right)^2 \ket{0^\text{e}}
\end{equation}
because all of the terms involving the photon creation / annihilation operators vanish once we take the expectation value and integrate out the photon degrees of freedom. Note also that the dipole self-energy term, in the Born-Oppenheimer approximation, only depends on the electronic part of the dipole operator, $\muElec$, because the nuclear part cancels once we take the expectation value.

From this point, one can easily adapt this polaritonic mean-field theory to obtain a QED generalization of Kohn-Sham DFT, or QED-DFT.\cite{DePrince22_9303,Rubio22_094101,DePrince23_5264} The QED-DFT ground-state is modeled by a non-interacting state of the form given above (Eq.~\ref{EQN:QED_HF}), with $|0^\text{e}\rangle$ now referring to a determinant of Kohn-Sham orbitals. The electronic part of the energy, $\bra{0^\text{e}} \hat{H}_\text{e} \ket{0^\text{e}}$, is handled as in standard Kohn-Sham DFT, {\em i.e.}, it is replaced by the Kohn-Sham energy $E_{\rm KS}$, which is a sum of core Hamiltonian ($h$) and classical Coulomb ($J$) contributions, plus an exchange-correlation functional of the electronic density ($\rho$), the gradient of the density ($\nabla \rho$), etc.:
\begin{equation}
E_{\rm KS} = \sum_{\mu\nu} ( h_{\mu\nu} + \frac{1}{2} J_{\mu\nu})\gamma_{\mu\nu} + f_{\rm xc}(\rho, \nabla \rho, ...)
\end{equation}
with
\begin{equation}
    J_{\mu\nu} = \sum_{\lambda \sigma} (\mu\nu|\lambda\sigma) \gamma_{\lambda \sigma} 
\end{equation}
Here, the Greek labels represent atomic basis functions,  $\gamma_{\mu\nu}$ is the one-particle density matrix, and $(\mu\nu|\lambda\sigma)$ is a two-electron repulsion integral in chemists' notation. 

The $\bm{\lambda}$-dependent part of the energy (the dipole-self energy) is handled the same way as it is treated in QED-HF, so the total QED-DFT energy is simply
\begin{equation}
\label{EQN:QED_DFT_ENERGY}
    E_{\rm QED-DFT} = E_{\rm KS} + \frac{1}{2} \bra{0^\text{e}} \left(\muDiffe\right)^2 \ket{0^\text{e}}
\end{equation}
More explicitly, the dipole self-energy term can be evaluated as
\begin{eqnarray}
\label{EQN:DSE_FINAL_FINAL}
    \frac{1}{2} \bra{0^\text{e}}  [{\bm{\lambda}} \cdot ({\muElec} - \langle {\muElec} \rangle)]^2 \ket{0^\text{e}} = \frac{1}{2} ({\bm{\lambda}}\cdot\langle {\muElec}\rangle)^2 \nonumber \\
    + \sum_{\mu\nu} (\frac{1}{2} J^{\rm DSE}_{\mu\nu} - \frac{1}{2} K^{\rm DSE}_{\mu\nu} + O^{\rm DSE}_{\mu\nu} ) \gamma_{\mu\nu} 
\end{eqnarray}
with
\begin{align}
\label{EQN:JDSE}
    J^{\rm DSE}_{\mu\nu} & = d_{\mu\nu} \sum_{\lambda \sigma} d_{\lambda \sigma} \gamma_{\lambda\sigma} =   ({\bm \lambda} \cdot \langle {\muElec}  \rangle) d_{\mu\nu} \\
\label{EQN:KDSE}
    K^{\rm DSE}_{\mu\nu} & = \sum_{\lambda \sigma} d_{\mu\sigma} d_{\lambda \nu} \gamma_{\lambda\sigma} \\
    O^{\rm DSE}_{\mu\nu} & = -( {\bm{\lambda}}\cdot \langle {\muElec}\rangle ) d_{\mu\nu} - \frac{1}{2} q_{\mu\nu}
\end{align}
Here, $\langle {\muElec}  \rangle$ represents the expectation value of the electronic dipole operator evaluated with respect to the Kohn-Sham state, and, following Ref.~\citenum{DePrince22_9303} , we have introduced 
\begin{equation}
    d_{\mu\nu} = \sum_{a \in \{x,y,z\}} \lambda_a \int \phi^*_\mu [-r_a] \phi_{\nu} d\tau,
\end{equation}
and
\begin{equation}
    q_{\mu\nu} = \sum_{ab \in \{x,y,z\}} \lambda_a \lambda_b \int \phi^*_\mu [-r_a r_b] \phi_{\nu} d\tau.
\end{equation}
which are ${\bm \lambda}$-weighted dipole and quadrupole integrals, respectively. The symbol $\phi_\mu$ represents an atomic basis function, $\lambda_a$ is a cartesian component of the coupling vector, ${\bm{\lambda}}$, and $r_x = x$, etc. Several terms in Eq.~\ref{EQN:DSE_FINAL_FINAL} cancel, so Eq.~\ref{EQN:QED_DFT_ENERGY} simplifies to
\begin{equation}
    \label{EQN:QED_DFT_ENERGY_FINAL}
    E_{\rm QED-DFT} = E_{\rm KS} - \frac{1}{2} \sum_{\mu\nu}( q_{\mu\nu} + K_{\mu\nu}^{\rm DSE}) \gamma_{\mu\nu}
\end{equation}

\subsection{Analytic Energy Gradients for QED-DFT}

When the QED-DFT problem is represented within the coherent-state basis, the gradient of the energy (Eq.~\ref{EQN:QED_DFT_ENERGY_FINAL}) with respect to a perturbation, $\chi$, is
\begin{equation}
    \frac{\partial E_{\rm QED-DFT}}{\partial \chi} = \frac{\partial E_{\rm KS}}{\partial \chi} - \frac{1}{2} \frac{\partial}{\partial \chi} \sum_{\mu\nu}( q_{\mu\nu} + K_{\mu\nu}^{\rm DSE}) \gamma_{\mu\nu}
\end{equation}
Here, $\frac{\partial E_{\rm KS}}{\partial \chi}$ equivalent to the gradient of the energy in standard (non-QED) Kohn-Sham DFT, so we focus on the the ${\bm \lambda}$-dependent term. We introduce ${\bm \lambda}$-weighted derivative dipole and {\color{black}quadrupole} integrals, defined as 
\begin{equation}
    d^\chi_{\mu\nu} = \frac{\partial}{\partial \chi} \sum_{a \in \{x,y,z\}} \lambda_a \int \phi^*_\mu [-r_a] \phi_{\nu} d\tau,
\end{equation}
and
\begin{equation}
    q^\chi_{\mu\nu} = \frac{\partial}{\partial \chi} \sum_{ab \in \{x,y,z\}} \lambda_a \lambda_b \int \phi^*_\mu  [-r_a r_b] \phi_{\nu} d\tau.
\end{equation}
respectively, and the final gradient expression is
\begin{equation}
    \frac{\partial E_{\rm QED-DFT}}{\partial \chi} = \frac{\partial E_{\rm KS}}{\partial \chi} - \frac{1}{2}\sum_{\mu\nu} q^\chi_{\mu\nu} \gamma_{\mu\nu} - \sum_{\mu\nu} \gamma_{\mu\nu} \sum_{\lambda \sigma} \gamma_{\lambda \sigma} d^\chi_{\mu \sigma} d_{\lambda \nu}
\end{equation}

\section{Computational Details}

\label{SEC:COMPUTATIONAL_DETAILS}

We have implemented the QED-DFT approach, together with QED-DFT analytic energy gradients in \texttt{hilbert},\cite{hilbert} which is a plugin to the \textsc{Psi4}\cite{Sherrill20_184108} electronic structure package. All cavity-free DFT {\color{black}and QED-DFT} calculations were performed using the aug-cc-pVDZ basis set and the dispersion-corrected B3LYP-D3BJ functional{\color{black}; a relatively large Lebedev-Laikov quadrature grid (with 540 spherical points) is used to ensure an accurate description of small energy changes associated with rotations with respect to the cavity mode polarization axis.} Following Ref.~\citenum{Rubio23_2766}, all calculations {\color{black}involving cavity interactions} use a cavity frequency ($\omega_\text{cav}$) and coupling strength ($\lambda$) equal to 1.5 eV and  0.1 a.u, respectively. Optimal geometries for the reactant (educt) and product species relevant to the Diels-Alder reactions we consider were {\color{black}obtained} using the \texttt{Pysisyphus}\cite{Grafe21_e26390} software suite, which we interfaced with \texttt{hilbert} to provide access to QED-DFT gradients. The calculations of cavity-free and cavity-embedded species were performed using DFT and QED-DFT, respectively. Initial estimates of transition state geometries were determined from the nudged elastic band (NEB) method\cite{Chakraborty04_7877, Zimmerman13_184102, Zheng16_094104} and further refined using the improved dimer method\cite{Keil05_224101} within \texttt{Pysisyphus}. {\color{black}We also performed QED-CC with single and double electronic excitations (QED-CCSD), plus the simultaneous creation of up to one photon (QED-CCSD-21) or up to two photons (QED-CCSD-22). All energy calculations with QED-CCSD-21 and QED-CCSD-22 were performed using the cc-pVDZ basis set. These methods have also been implemented in \texttt{hilbert}.}

The QED-DFT energy depends on the relative orientation of the molecular species and the cavity-mode polarization axis. For this reason, a traditional internal coordinate system is a poor choice {\color{black}when performing geometry optimizations} in that it overlooks the spatial orientation of the system{\color{black}, and} procedures for the back-transformation to cartesian coordinates can be ambiguous. {\color{black} This ambiguity can be removed by performing the back transformation in a way that minimizes the deviation between the atomic positions for the current structure and the structure from the preceding optimization step, but such a procedure could counteract cavity-induced rotational forces}. As such, all geometry optimizations in this work were performed using translation-rotation internal coordinates\cite{Song16_214108} (TRIC),  {\color{black} which incorporate rotational degrees of freedom in the coordinate system that permit the molecule to reorient in space. A cartesian coordinate system could also suffice, but the TRIC system has fewer degrees of freedom. The \texttt{Pysisyphus} infrastructure supports the TRIC system.}

\section{Results and Discussion}

\label{SEC:RESULTS}
\newcommand{\paren}[1]{\left( #1 \right)}

\begin{figure*}[htbp]
    \centering
    \includegraphics[width=\linewidth]{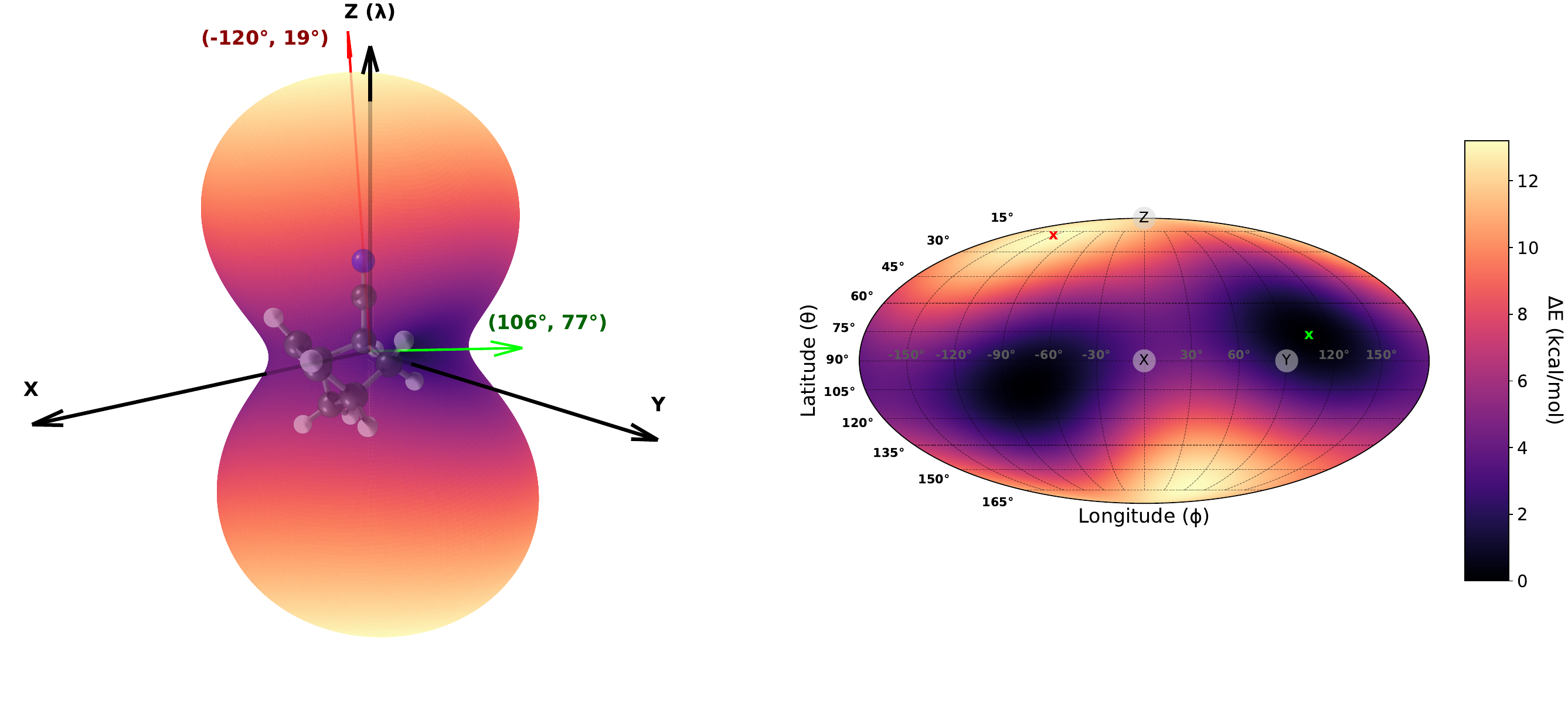}
    \caption{
    The energy profile of the $endo$ Diels-Alder product of cyclopentadiene and acrylonitrile as a function of the spatial orientation of the molecular dipole moment for the cavity-free geometry with respect to a $z$-polarized cavity. The $endo$ product is visualized in (a), with the dipole oriented along the $z$-axis. The energy profile is represented in two ways: (a) as a deformed sphere where the radius and color at each point represent the relative energy difference from the minimum energy orientation, and (b) as a Mollweide projection{\color{black}. In (a), the minimum energy orientation (MEO) and highest energy orientation (HEO) of the dipole moment are denoted by green and red arrows, respectively, and the coordinates in parenthesis are ($\phi$, $\theta$), or the azimuthal and polar angles, respectively. In (b), the MEO and HEO are denoted by green and red x's, respectively.} } 
    \label{fig:sphere_energy}
\end{figure*}
In gas- or solution-phase small-molecule reactions, it is reasonable to expect that the molecular species are free to rotate with respect to a fixed coordinate frame. Without an explicit mechanism to fix the molecular orientation with respect to the cavity-mode polarization, it is important to understand the rotational energy landscape and its impact on the optimal reaction pathway and any inferred properties. We emphasize this importance by revisiting the study from Ref.~\citenum{Rubio23_2766} that explored the thermodynamics of a Diels-Alder cycloaddition involving cyclopentadiene and acrylonitrile in an optical cavity.

\subsection{Rotational energy landscape}
\label{SEC:energy_landscape}

Figure~\ref{fig:sphere_energy} illustrates the energy of the $endo$ Diels-Alder product of cyclopentadiene and acrylonitrile (see Fig.~\ref{fig:reaction_diagram}) as a function of the relative orientation of the molecular dipole moment and the cavity mode polarization axis, which is aligned along the $z$-direction.  In this example, the internal coordinates of the molecule are frozen at the optimal values from a geometry optimization performed in the absence of the cavity.  The surface on the left-hand side of Fig.~\ref{fig:sphere_energy} represents the energy of the system as a function of the direction of the total dipole moment, which is characterized by the {\color{black} azimuthal} ($\phi$) and {\color{black} polar} ($\theta$) angles relative to the cavity mode polarization axis (the $z$-axis). The radius of the surface ($\rho$) is proportional to the difference between the energy at $\phi$ and $\theta$ and the energy at the minimum energy orientation (MEO) at $\phi = \phi^\prime$ and $\theta = \theta^\prime$, {\em i.e.},
\begin{equation}
    \label{eq:rho}
    \rho\paren{\phi,\theta} = E\paren{\phi,\theta | {\bf R}} - E\paren{\phi^\prime,\theta^\prime | {\bf R}} + \rho_0
\end{equation}
where ${\bf R}$ represents the fixed internal coordinates of the cavity-free $endo$ Diels-Alder product. The radial function is given an arbitrary minimum length ($\rho_0=0.10$~kcal mol$^{-1}$) so that energy changes appear as deformations to a small sphere. The color of the surface also reflects the difference in energies evaluated at a given orientation and at the MEO. The color gradient is defined on the right-hand side of the figure, where the energy landscape is also presented as a heat map.

The MEO corresponds to {\color{black} $\phi^\prime=106\degree$ and $\theta^\prime=77\degree$}, where the molecular dipole is oriented along the $xy$-plane.
The energy increases rapidly as the dipole orientation approaches the $z$-axis (${\color{black} \theta} \to 0\degree$ or $180\degree$) for any {\color{black} $\phi$ value}. To understand this behavior, consider the form of the QED-DFT energy, when the Hamiltonian is represented within the coherent-state basis. Equation \ref{EQN:QED_DFT_ENERGY} suggests that the system will be most stable when the molecule is oriented such that the dipole self-energy is minimized. If the cavity mode is polarized along the $z$ direction [{\em i.e.}, ${\bm \lambda} = (0, 0, \lambda)]$, the dipole self-energy, $E_\text{DSE}$, is
\begin{equation}
    \begin{aligned}
        E_\text{DSE} &= \frac{1}{2} \bra{0^{\rm e}} \left(\muDiff\right)^2 \ket{0^{\rm e}}\\
        &= \frac{1}{2}\lambda^2 \bra{0^{\rm e}} \left ( \hat{\mu}^z_\text{e} - \langle \hat{\mu}^z_\text{e} \rangle  \right)^2 \ket{0^{\rm e}}
    \end{aligned}    
\end{equation}
Hence, the key quantity for understanding the rotational energy landscape of the cavity-bound species is the variance in the {\color{black}electronic} dipole moment along the $z$ direction. Indeed, the dipole variance along $z$ is minimized at the MEO, and maximized when {\color{black} $\phi=-120\degree$ and $\theta=19\degree$} at the highest energy orientation (HEO) of the molecule. 

The HEO lies higher in energy than the MEO by $13.2$~kcal/mol. This gap is substantial and comparable to the $13.5$~kcal/mol cavity-free activation energy for the Diels-Alder reaction that produces this product. These results have significant implications for the thermodynamics and kinetics of chemical reactions carried out in optical cavities. In the absence of a mechanism to restrict the rotational degrees of freedom, a cavity-embedded molecule should reorient to minimize the dipole self-energy. Conclusions derived from energetic analyses carried out at fixed molecular orientations miss this important detail.

\subsection{Challenges when optimizing non-bonded complexes}

\label{SEC:CHALLENGES}

\begin{figure*}[!hp]
    \centering
    \includegraphics[height=0.65\paperheight]{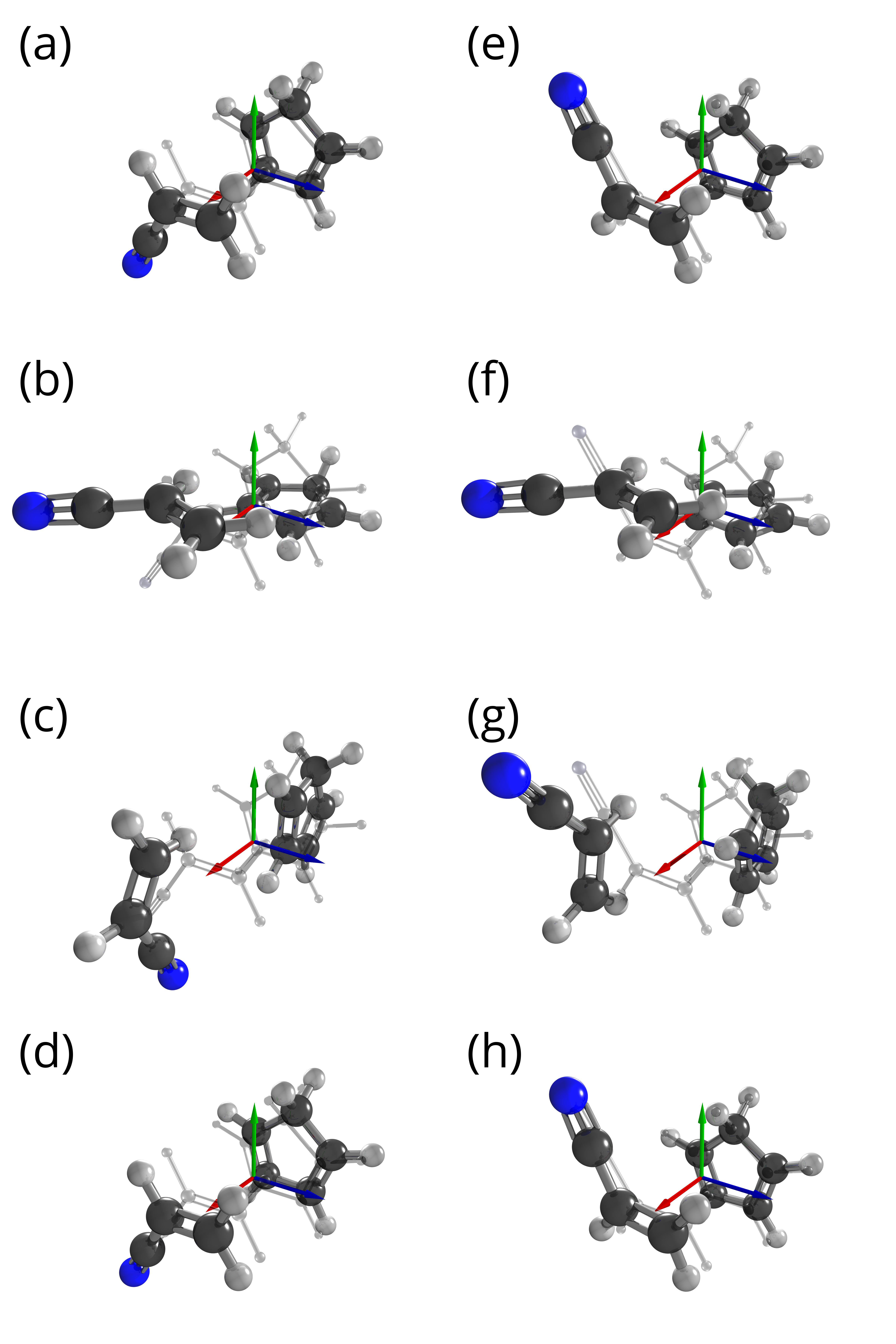}
    \caption{Educt structures for the $endo$/$exo$ pathways optimized under different cavity mode polarizations.
        The {\color{black} left (a-d) and right (f-h)} columns correspond to the $endo$ and $exo$ geometries.
        {\color{black}QED-DFT geometries obtained from optimizations starting from $x$-, $y$-, and $z$-polarized cavity modes [panels (a,e); (b,f); (c,g), respectively], and QED-DFT geometries obtained from the optimization protocol developed in this section, starting from an $x$-polarized cavity mode (d,h). 
        A partially transparent representation of the molecular geometry optimized in the absence of the cavity is included in each panel.}
        }
    \label{fig:educt_geometries}
\end{figure*}

As mentioned in Sec.~\ref{SEC:COMPUTATIONAL_DETAILS}, {\em ab initio} QED geometry optimizations should be performed using a coordinate system that captures the orientation dependence of the energy; we use the TRIC system. This choice alone does not completely eliminate the numerical difficulties in optimizing geometries of cavity-embedded species, particularly for non-bonded complexes. For example, in the Diels-Alder educt structure, acrylonitrile has a larger dipole variance than cyclopentadiene, so the former fragment may experience larger forces due to the dipole self-energy contribution to the gradient. As a result, situations arise where the system moves away from the desired geometry in the course of an optimization. Figure \ref{fig:educt_geometries} illustrates the educt structures for the $endo$ and $exo$ pathways (left and right panels, respectively) that were optimized with the system interacting with a cavity mode polarized along different cartesian axes. The axes are chosen to be the same as those in Ref.~\citenum{Rubio23_2766}, where the $x$-axis is along the forming carbon-carbon bond, and the $yz$ plane is along the plane of the cyclopentadiene such that the $xy$-plane intersects its sp$^3$ hybridized carbon and the $xz$ plane intersects the adjacent carbon atoms. {\color{black}The structures in panels (a,e), (b,f), and (c,g) were optimized with the cavity mode polarized along the $x$-, $y$-, and $z$-direction, respectively. In all panels, the optimized structures are overlaid on partially transparent representations of the initial structures, which were optimized at the B3LYP-D3BJ/aug-cc-pVDZ level of theory in the absence of the cavity.  It is clear that geometry optimizations performed in the presence of cavity interactions may cause the molecular fragments to move away from their initial stacked geometries, leading to stationary points that are not relevant for the cyclization reaction in question. For example, the $y$- and $z$-polarized cavity modes cause the fragments to align side by side, while the $x$-polarized cavity modes preserve the stacked orientation. Surprisingly, despite the clear structural differences, the configurations represented in panels (a-c) and (e-g) lie within 0.3 kcal mol$^{-1}$ of each other.} 

The large variations in the optimized structures depicted in Fig.~\ref{fig:educt_geometries}, some of which are qualitatively different than the educt structures relevant to the cyclization reaction, are problematic and motivate the need for a robust protocol for geometry optimizations in the presence of an optical cavity. As such, we have developed and applied the following procedure. An initial constrained optimization is performed such that the entire complex can rotate into its optimal orientation relative to the cavity mode polarization axis. For the educt structures relevant to the Diels-Alder reactions considered in this work, the relative orientations of the fragments are partially fixed by constraining the distance between the bond-forming carbon atoms, as well as the dihedral angles between fragment planes. Once the optimal orientation of the constrained educt structure has been identified, a full geometry optimization can be performed. With this procedure, the fully optimized educt structures associated with the $endo$ path differ in energy by only $\approx$ 0.01 kcal mol$^{-1}$, while those for the $exo$ path differ by $\approx$ 0.03--0.04 kcal mol$^{-1}$, regardless of the initial relative orientation of the complex and the cavity mode polarization axis. {\color{black} Panels (d) and (h) of Fig.~\ref{fig:educt_geometries} depict structures optimized according to this procedure, starting from an $x$-polarized cavity mode. Unlike some of the structures obtained from the naive geometry optimization procedure, these structures, as well as those optimized starting from $y$- and $z$-polarized cavity modes (not shown), retain the desired relative orientations of the fragments for the cyclization reaction.} 

{\color{black}Before moving on, we briefly comment on general structural changes that are observed for the non-bonded educt species when they are allowed to relax in the presence of the cavity, using the protocol developed in this section [panels (d) and (h) of of Fig.~\ref{fig:educt_geometries}]. First, cavity interactions lead to appreciable changes to the inter-fragment distances (measured by the distance between the centroids of the two fragments); the inter-fragment distances in the $endo$ and $exo$ educt structures increase by $0.99$~\AA~and $0.87$~\AA, respectively. Second, we can quantify orientational changes using the angles ($\phi$, $\theta$) that characterize the orientation of the dipole moment, as discussed in Sec.~\ref{SEC:energy_landscape}. The dipole moments are initially aligned at ($\phi=101\degree$, $\theta=55\degree$) and  ($\phi=-101\degree$, $\theta=47\degree$) for the $endo$ and $exo$ educt structures, respectively. After relaxation, these angles change by ($\Delta\phi=-10\degree$, $\Delta\theta=3\degree$) and ($\Delta\phi=10\degree$, $\Delta\theta=6\degree$), respectively. These changes correspond to the rotation required for the molecule to adopt its relaxed minimum energy orientation.}

\subsection{Enthalpies and barrier heights for the Diels-Alder reaction}

\begin{table*}[htbp!]
{\color{black}

    \centering
    \setlength{\tabcolsep}{7pt}
    \begin{tabular}{clrr|rr|rr}
        \toprule
        \multirow{2}{*}{path} & \multirow{2}{*}{structures} & \multicolumn{2}{c|}{\footnotesize QED-DFT} & \multicolumn{2}{c|}{\footnotesize ${}^*$QED-CCSD-21} & \multicolumn{2}{c}{\footnotesize ${}^*$QED-CCSD-22} \\
        \cline{3-8}
        && \multicolumn{1}{c}{$H$} & \multicolumn{1}{c|}{$E_a$} & \multicolumn{1}{c}{$H$} & \multicolumn{1}{c|}{$E_a$} & \multicolumn{1}{c}{$H$} & \multicolumn{1}{c}{$E_a$} \\
        \midrule
        && \multicolumn{6}{c}{$\lambda = 0.00$} \\
        \cline{3-8}\\[-.5em]
        $endo$       & \multirow{4}{*}{fixed}   & -22.5 & 13.5 & -30.1 & 21.7 & -30.1 & 21.7 \\
        $exo$        &                          & -22.3 & 13.8 & -29.7 & 21.3 & -29.7 & 21.3 \\\\[-1em]
        $endo - exo$ &                          &  -0.2 & -0.3 &  -0.4 &  0.4 &  -0.4 &  0.4 \\
                                                                                       \\[-1em]
        
        && \multicolumn{6}{c}{$\lambda = 0.10$} \\
        \cline{3-8}\\[-.5em]
        \multirow{4}{*}{$endo$}        & relaxed     &  -9.3 & 34.6 & -27.7 & 26.2 & -27.9 & 25.9 \\
                                       & fixed ($x$) & -10.1 & 34.0 & -26.4 & 28.2 & -26.6 & 27.8 \\
                                       & fixed ($y$) & -37.6 &  7.8 & -35.8 & 18.5 & -35.6 & 18.6 \\
                                       & fixed ($z$) & -22.6 & 20.1 & -31.2 & 22.7 & -31.2 & 22.7 \\[1em]
        \multirow{4}{*}{$exo$}         & relaxed     & -11.8 & 30.7 & -29.2 & 24.9 & -29.4 & 24.7 \\
                                       & fixed ($x$) & -10.4 & 33.8 & -27.1 & 27.7 & -27.3 & 27.3 \\
                                       & fixed ($y$) & -35.6 & 10.0 & -35.6 & 18.9 & -35.4 & 19.0 \\
                                       & fixed ($z$) & -23.2 & 20.2 & -32.1 & 22.3 & -32.0 & 22.2 \\\\[-1em]
        \multirow{4}{*}{$endo - exo$}  & relaxed     &   2.5 &  3.9 &   1.5 &  1.3 &   1.5 &  1.2 \\
                                       & fixed ($x$) &   0.3 &  0.2 &   0.7 &  0.5 &   0.7 &  0.5 \\
                                       & fixed ($y$) &  -2.0 & -2.2 &  -0.2 & -0.4 &  -0.2 & -0.4 \\
                                       & fixed ($z$) &   0.6 & -0.1 &   0.9 &  0.4 &   0.8 &  0.5 \\                                
      \bottomrule
    \end{tabular}

    \caption{Reaction enthalpies and activation energies (kcal mol$^{-1}$) corresponding to the $endo$ and $exo$ pathways for different structures (optimized/fixed), cavity mode polarizations ($x$,$y$,$z$), and levels of theory.\\    
    $\null^*$ evaluated with the cc-pVDZ basis
    } 
    \label{tab:thermotable}
}
\end{table*}

We now turn our attention to the the cavity-induced changes to the thermodynamics of the Diels-Alder reactions depicted in Fig.~\ref{fig:reaction_diagram}. {\color{black}Table~\ref{tab:thermotable} contains reaction enthalpies ($H$) {\color{black} and activation energies ($E_a$)} for the $endo$ and $exo$ pathways of the Diels-Alder reaction in the absence of the cavity ($\lambda$ = 0.0 a.u.) and at $\lambda$ = 0.1 a.u. We consider four sets of calculations on cavity-embedded species: one set where the structures are fully optimized at the QED-B3LYP-D3BJ/aug-cc-pVDZ level of theory {(labeled ``relaxed'') and three sets where we use fixed structures optimized in the absence of the cavity and fixed orientations relative to different cavity mode polarization axes [labeled ``fixed ($x$)'', etc.].
Given either ``fixed'' or ``relaxed''} educt and product geometries, transition states are located} using the NEB implementation in \texttt{Pysisyphus}, which can capture the cavity-induced rotation of the molecule as the cycloaddition progresses. This rotation is necessary to account for the different orientation preferences of the optimized educt and product structures. The transition state structures are further refined using the improved dimer method\cite{Keil05_224101} within \texttt{Pysisyphus}.

In the cavity-free case, the $endo$ and $exo$ pathways {\color{black}exhibit comparable energetics; the enthalpies and activation energies differ by less than 1 kcal mol$^{-1}$ at both the DFT and CCSD levels of theory.} The question so-often posed in {\em ab initio} cavity QED studies is then, can ultra-strong light-matter coupling induce meaningful changes to the enthalpy or barrier height for these reactions? The answer to this question is, of course, yes, but wildly different conclusions can be drawn depending on whether the structures are relaxed in the presence of the cavity or, with fixed structures, how the molecule is oriented with respect to the cavity mode polarization axis. {\color{black}Consider the QED-DFT description of the} fully optimized structures, {\color{black}where} we observe the following. First, the enthalpy of the reaction is reduced in magnitude by {\color{black}roughly} a factor of two, for both pathways. Second, the thermodynamic favorability of the {\color{black}$exo$ pathway is enhanced, relative to that of the $endo$ pathway, with the enthalpy being lower by roughly by 2.5 kcal mol$^{-1}$} in the former case. Third, the barrier to the forward reaction increases by more than a factor of two for both path{\color{black}ways}, {\color{black}with the barrier height for the $endo$ pathway being higher, by nearly 4 kcal mol$^{-1}$}. These results suggest that ultrastrong light-matter coupling could be used to alter the preferred cycloaddition product, with the caveat that the forward reaction would be much less {\color{black}kinetically} favorable for the cavity embedded system, relative to the original cavity-free system. 

The story {\color{black}can be} quite different when considering unrelaxed geometries with fixed orientations relative to the cavity mode polarization axis. For both the $endo$ and $exo$ pathways, we find that an $x$-polarized cavity mode leads to similar results as with fully relaxed structures; the magnitudes of the enthalpies are reduced by more than a factor of two, and the barrier heights more than double. On the other hand, $y$- and $z$-polarized cavity modes lead to significantly more negative enthalpies, while changing the forward reaction barrier heights by only 1--2 kcal mol$^{-1}$.
{\color{black} 
For fixed geometries relative to a $y$-polarized cavity mode, the $endo$ pathway is thermodynamically favored by 2.0 kcal mol$^{-1}$, while $x$- and $z$-polarized cavity modes result in a slight preference for the $exo$ pathway, by only 0.3 and 0.6  kcal mol$^{-1}$, respectively. We stress that these results are at odds with results of calculations performed on fully-optimized structures, which instead predict that the $exo$ product is clearly the preferred thermodynamic product, by 2.5 kcal mol$^{-1}$.}

{\color{black}

To this point, we have focused on QED-DFT-based descriptions of cavity-modified chemistry, which lack explicit electron-photon correlation effects. To gain an understanding of the role of these effects and also to provide a more direct comparison with the results from Ref.~\citenum{Rubio23_2766}, we have performed QED-CC calculations that consider single and double electronic excitations (QED-CCSD), plus the creation of up to one or two photons (QED-CCSD-21 or QED-CCSD-22, respectively). The results of these calculations, carried out using DFT/QED-DFT-optimized geometries and the cc-pVDZ basis set, are provided in Table~\ref{tab:thermotable}. Note that our QED-CCSD-21 results for fixed geometries are slightly different than those of Ref.~\citenum{Rubio23_2766} because our fixed geometries are derived from DFT/aug-cc-pVDZ  calculations, whereas Ref.~\citenum{Rubio23_2766} considered geometries optimized at the MP2/cc-pVDZ level of theory. The differences in barrier heights and enthalpies presented in Table \ref{tab:thermotable} and those provided in Ref.~\citenum{Rubio23_2766} differ by at most 2 kcal mol$^{-1}$, and we stress that these differences are entirely structural in origin. We have verified the equivalence of our QED-CCSD-21 and QED-CCSD-22 implementations and those of Ref.~\citenum{Rubio23_2766} by comparing energies evaluated using the MP2/cc-pVDZ-optimized geometries from that work to the results provided in the associated supporting information.

As can be seen in Table \ref{tab:thermotable}, QED-CCSD results obtained when considering the creation of up to one photon (QED-CCSD-21) or up to two photons (QED-CCSD-22) are quite similar. Enthalpies and barrier heights evaluated using these two methods differ by at most only 0.4 kcal mol$^{-1}$. Some of the qualitative observations drawn from QED-DFT calculations apply to the QED-CCSD case. First, we observe a decrease in the reaction enthalpy and an increase in the forward reaction barrier height for both the $endo$ and $exo$ pathways when the molecule is coupled to the cavity, for relaxed structures. However, the magnitudes of these changes are much less than in the case of QED-DFT. Second, as was found for QED-DFT above and for QED-CCSD in Ref.~\citenum{Rubio23_2766}, cavity-coupled barrier heights are larger, compared to cavity-free values, for $x$- and $z$-polarized cavity modes, and smaller for the $y$-polarized mode. Similarly, as was found for QED-DFT and in Ref.~\citenum{Rubio23_2766}, reaction enthalpies are more negative, compared to cavity-free values, for $y$- and $z$-polarized cavity modes, and less negative for the $x$-polarized mode. Again, the magnitudes of the changes induced by the cavity are somewhat smaller for QED-CCSD than for QED-DFT. Finally, like QED-DFT, QED-CCSD calculations at $lambda = 0.1$ a.u.~on relaxed geometries predict that the $exo$ product is clearly thermodynamicaly preferred one (by 1.5 kcal mol$^{-1}$). On the other hand, when using fixed geometries, the thermodynamic preference for the $exo$ or $endo$ pathway changes depending on the cavity mode polarization, but in any case, the enthalpies and barrier heights for the two pathways differ by less than 1 kcal mol$^{-1}$. 

}

\section{Conclusions}

\label{SEC:CONCLUSIONS}

A number of recent theoretical studies have focused on the application of {\em ab initio} cavity QED methodologies to the ground-state electronic structure of cavity embedded species, in an effort to identify cases where ultra-strong light-matter coupling leads to useful or interesting changes to chemistry. The common theme in these studies is that non-trivial changes ({\em e.g.}, to reaction enthalpies, barrier heights, etc.) can be realized, given large enough light-matter coupling strengths, but these changes tend to display a pronounced orientational dependence. None of the studies cited in Sec.~\ref{SEC:INTRODUCTION} consider geometry relaxation or rotational effects that should be induced by the very large coupling strengths employed in those works. 

In this study, we have shown that geometry relaxation effects can lead one to draw qualitatively different conclusions regarding the impact of an optical cavity on ground-state chemistry, as compared to calculations involving fixed structures and orientations relative to the cavity mode axis. We have demonstrated this point by revisiting the Diels-Alder cycloaddition reactions considered in Ref.~\citenum{Rubio23_2766}. Without a physical mechanism to constrain the orientation of the molecule in the cavity, the reactant / transition state / product species reorient in a way that eliminates the dependence on the initial relative orientations of the molecular components and the cavity mode polarization axis. Our calculations involving fully-relaxed structures suggest that the $exo$ cycloaddition product is {\color{black}clearly} the thermodynamically favored one at large coupling strengths, while calculations on fixed structures {\color{black}can} lead to {\color{black}different} interpretations{\color{black}; depending on the chosen polarization axis, the preference for the $exo$ pathway can be significantly reduced, or the $endo$ product can become the thermodynamically preferred one}. As such, conclusions drawn from energetic analyses based on {\em ab initio} QED calculations that use fixed molecular structures should be viewed with caution.

{\color{black}Lastly, we note that few groups have reported analytic energy gradients for QED generalizations of electronic structure methods. Aside from the present work, Shao and coworkers have developed analytic energy gradients for QED generalization of time-dependent DFT,\cite{Shao22_124104} and, more recently, Lexander, Angelico,  Kj{\o}nstad, and Koch presented analytic energy gradients for QED-CC theory.\cite{Koch24_2406.08107} In the absence of analytic gradients, numerical gradients could prove too computationally demanding to be practical for some applications. In such cases, cavity-induced rotational effects in the ground state could be taken into account simply by orienting the system such that the dipole self-energy is minimized; the dominant effects could even be estimated from mean-field calculations performed in the absence of any cavity interactions. Alternatively, one could use orientational averaging techniques, such as those described in Refs.~\citenum{Koch24_e1684} and \citenum{Koch23_2308.06181}. }

\vspace{0.5cm}

{\bf Supporting Information} 
Relaxed and unrelaxed geometries for the  \textit{endo} and \textit{exo} Diel-Alder reactions at the B3LYP-D3BJ/aug-cc-pVDZ level of theory.

\begin{acknowledgments}This material is based upon work supported by the National Science Foundation under Grant No. CHE-2100984. \\ 
\end{acknowledgments}

\noindent {\bf DATA AVAILABILITY}\\

    The data that support the findings of this study are available from the corresponding author upon reasonable request.

\bibliography{main}

\end{document}